\documentclass[a4paper,12pt]{article}

\input amssym.def
\input amssym.tex

\def\Hz{\Bbb{H}}
\def\Rz{\Bbb{R}}

\def\beq{\begin{equation}} 
\def\eeq{\end{equation}} 
\def\lbeq#1{\begin{equation} \label{#1}} 
\def\gzit#1{{\rm (\ref{#1})}} % (e. g. equation (7)
\def\fct#1{\mathop{\rm #1}}% ==> \sin is equivalent to \fct{sin}

\def\hbar{h\hspace{-2mm}^-}

\def\<{\langle} 		% expectation
\def\>{\rangle} 		% expectation
\def\tr{\fct{tr}~}

\def\half{\frac{1}{2}} 
\def\eps{\varepsilon}
\def\wave{\raisebox{-0.6ex}{\symbol{126}}}

\begin{document}

\vspace*{-2cm}

\begin{center}

{\LARGE \bf On a realistic interpretation of quantum mechanics} 

\vspace{1cm}

{\large \bf Arnold Neumaier} \\
~\\
Institut f\"ur Mathematik\\
Universit\"at Wien\\
Strudlhofgasse 4\\
A-1090 Wien, Austria\\
email: neum@cma.univie.ac.at\\
WWW: http://solon.cma.univie.ac.at/\wave neum/\\
~\\

\end{center}

{\bf Abstract.} The best mathematical arguments against a realistic 
interpretation of quantum mechanics -- that gives definite but 
partially unknown values to all observables -- are analysed and shown 
to be based on reasoning that is not compelling. 

This opens the door for an interpretation that, while respecting 
the indeterministic nature of quantum mechanics, allows to speak of 
definite values for all observables at any time that are, however, 
only partially measurable. 

The analysis also suggests new ways to test the foundations of  
quantum theory.

\begin{flushleft}

\vspace{0.5cm}

{\bf 1991\hspace{.4em} MSC Classification}: 
             primary 81P10\\
\vspace{0.5cm}

{\bf 1990\hspace{.4em} PACS Classification}: 
             03.65\\
\end{flushleft}

\newpage 

% \bigskip
%%%%%%%%%%%%%%%%%%%%%%%%%%%%%%%%%%%%%%%%%%%%%%%%%%%%%%%%%%%%%%%%%%%%%%%%
\section{Introduction} \label{intro}

Quantum physics is a very successful theory for predicting nature.
However, in spite of many attempts, a mathematically and philosophically
convincing basis for the interpretation of quantum phenomena has not 
yet been found. 

As stated recently by {\sc Zeilinger} \cite{Zei},
there are at least two levels of interpreting quantum mechanics: the 
statistical interpretation in the narrower sense introduced by 
{\sc Born} \cite{Bor}, on which there is almost complete consensus 
between physicists, and the interpretation of the meaning of the 
quantum mechanical concepts, where no agreement has been reached and 
all existing interpretations have been found wanting.

The `orthodox' Copenhagen interpretation in terms of state reduction
by external measurement loses credit because it becomes meaningless for
the universe as a whole. There have been a number of proposals
to change the structure of quantum mechanics, e.g., through pilot 
waves ({\sc Bohm} \cite{Bohm}), consistent histories 
(e.g., {\sc Omn{\`e}s} \cite{Omn}), event enhancement 
({\sc Blanchard \& Jadczyk} \cite{BlaJ}), or gravitational objective 
reduction ({\sc Penrose} \cite{Pen}). 

\bigskip
Perhaps the main reason why there is so little progress on the meaning 
of quantum mechanical concepts is that it is usually seen as thoroughly enmeshed with measurement problems. I believe that this is a mistake,
and that more clarity can be obtained by separating the analysis of
meaning from that of observability. The discussion thereby becomes 
more concise and clear. 

Here I refer to {\em meaning} as a conceptual, logically consistent 
mathematical framework that allows one to speak unambiguously about 
{\em all} the terms used in the theory, in a way intuitively related 
to corresponding concepts of external reality. This part is often 
elegant and concise.

On the other hand, I refer to {\em observability} questions (or 
{\em measurement} problems) as the operational 
explanation of how to obtain quantitative values for the observables 
of a system, preferably within the framework of a  well-defined 
mathematical theory. This part is usually messy, and it is the part
that I propose to avoid until the other part is satisfactory. (This
also saves me from entering a discussion about the meaning of words
like `operational' or `obtain'.)

For example, in classical real analysis, infinitesimals are an 
ill-defined, meaningless concept, though they can be approximately 
realized, while random sequences are a well-defined, meaningful concept,
though there is no constructive way of finding one, except 
approximately.

\bigskip
A historical case supporting the power of this view of separating the 
analysis of meaning from that of observability is the uncertainty 
principle. It was argued by {\sc Heisenberg} \cite{Hei} in terms of 
observability, and later by {\sc Robertson} \cite{Rob} in mathematical 
terms as a simple consequence of the canonical commutation relation 
$[p,q]=-i\hbar$. The mathematical 
argument takes only a few lines and is simple and compelling; the 
discussion of observability is complex but shows that the results of 
the mathematics cannot be in conflict with 
what can in principle be measured by experiment. 

Further support for the positive effect of the separation of meaning 
from observability is given by Bell's inequality 
({\sc Bell} \cite{Bel}, {\sc Clauser} et al. \cite{ClaHS}), a purely 
algebraic statement whose clarity is compelling, and the subsequent 
verification of its violation through experiments by 
{\sc Aspect} \cite{Asp}. 

Finally, it seems that measurement problems can be adequately 
analysed by generalized observables defined as positive operator valued 
(POV) measures; see, e.g., {\sc Davies} \cite{Dav}, 
{\sc Busch} et al. \cite{BusGL}. The gradual and approximate state 
reduction can be explained thermodynamically through dissipative 
interaction by the coupling to a macroscopic apparatus or heat 
bath; see, e.g., {\sc Zurek} \cite{Zur}, {\sc Joos \& Zeh} \cite{JooZ},
{\sc Ghirardi, Rimini \& Weber} \cite{GhiRW}. 

That obtaining information about a system and state reduction 
are not equivalent things is a consequence of the possibility of 
nondemolition measurements ({\sc Braginsky} et al. \cite{BraVT}, 
{\sc Kwiat} et al. \cite{KwiWH}) that produce knowledge about a system 
but (almost) avoid state reduction.
And since the measuring apparatus necessarily 
involves a huge number of particles, any satisfactory solution of
measurement problems should not be a part of the foundational 
concepts and these but be obtained as their {\em consequence}. 

\bigskip
In an attempt to separate the discussion about the meaning of quantum 
mechanics from that of experimental consistency, I searched the 
literature for concise statements of the basic difficulties in terms
unrelated to measurement, but given in a clear mathematical context 
that allows one to analyse the logic of the problems without losing 
oneself in philosophical speculations.

The outcome was somewhat surprising: The best presented mathematical 
arguments turned out to have subtle flaws. This becomes apparent when 
one reduces them to arguments residing completely within the universally
accepted statistical interpretation in the narrower sense. 

Probably the weaknesses in the arguments escaped previous attention 
since the arguments were only used for philosophical discussion and 
justification of foundations, and never subjected to experiment. 
I therefore suggest below some specific things that one should try to
test with suitable experiments.

\bigskip
In the following, I shall present my analysis of some arguments 
related to a realistic interpretation of quantum mechanics; 
{\em realistic} in the sense that it allows to speak of definite values 
for {\em all} observables at any time that are, however, only partially 
measurable. 

The findings give rise to the hope that a consistent interpretation is 
possible that respects the stochastic nature of quantum mechanics,
but is also realistic, with all the accompanying advantages for our 
intuition. In a companion paper ({\sc Neumaier} \cite{Neu1}), 
a proposal for such a realistic interpretation will be presented.

\bigskip
%%%%%%%%%%%%%%%%%%%%%%%%%%%%%%%%%%%%%%%%%%%%%%%%%%%%%%%%%%%%%%%%%%%%%%%%
\section{Feynman's argument} \label{feynman}

In his 1948 paper ``Space-time approach to non-relativistic quantum 
mechanics'' \cite{Fey}, Richard {\sc Feynman} gives on pp.368-369 the 
following argument. 

{\em
``We define $P_{ab}$ as the probability that if the measurement $A$ gave
the result $a$, then measurement $B$ will give the result $b$. [...] 
denote by $P_{abc}$ the probability [...] if $A$ gives $a$, then $B$ 
gives $b$, and $C$ gives $c$. [...] If the events between $a$ and $b$ 
are independent of those between $b$ and $c$, then 
\lbeq{e1}
P_{abc}=P_{ab}P_{bc}. 
\eeq
This is true according to quantum mechanics
when the statement that $B$ is $b$ is a complete specification of the 
state. In any event, we expect the relation 
\lbeq{e2}
P_{ac}=\sum_b P_{abc}. 
\eeq
[...] Now, the essential difference between classical and quantum 
physics lies in Eq. \gzit{e2}. In classical 
mechanics it is always true. In quantum mechanics it is often false. 
[...] The classical law, obtained by combining \gzit{e1} and \gzit{e2},
\setcounter{equation}{3} 
\lbeq{e4}
P_{ac}=\sum_b P_{ab}P_{bc}
\eeq
is replaced by 
\lbeq{e5}
\phi_{ac}=\sum_b \phi_{ab}\phi_{bc}.
\eeq
[for probability amplitudes $\phi_{ab}$ with $|\phi_{ab}|^2=P_{ab}$]
If \gzit{e5} is correct, ordinarily \gzit{e4} is incorrect. [...]
Looking at probability from a frequency point of view \gzit{e4} simply 
results from the statement that in each experiment giving $a$ and $c$, 
$B$ had some value. [...] Noting that gzit{e5} replaces \gzit{e4} only 
under the circumstance that we make no attempt to measure $B$, we are 
lead to say that the statement, `$B$ had some value,' may be 
meaningless whenever we make no attempt to measure $B$.''
}

Feynman's argument sounds convincing but it has a flaw in the assumption
of \gzit{e1}, the conditional independence of $c$ from $a$, given $b$,
for all states summed over in \gzit{e2}. This assumption is 
inconsistent with quantum mechanics.

Indeed, one of the characteristic features of quantum mechanics is its
inseparability, the impossibility of splitting a quantum mechanical
system into two that are independent. According to the Copenhagen 
interpretation ({\sc Bohr} \cite{Boh}), and verified abundantly by 
experiment, the {\em whole} space-time set-up must be specified to 
determine the correct outcome of an experiment. The contradiction 
obtained by Feynman in comparing \gzit{e4} and 
\gzit{e5} therefore need not have the consequence
that $B$ sometimes has no value, but serves more naturally as 
{\em a strong argument for quantum inseparability}.

The Markov property \gzit{e1} does typically not even hold for 
consecutive events; due to the inherent noncommutativities, states 
"remember" something about their whole history and "anticipate" 
something about their whole future. To see this, pick $a,b,c$ as 
observations of the same observable at increasing times (for memory) 
and at decreasing times (for anticipation). It should be an interesting 
project to find the conditions under which (and the accuracy to which) 
the Markov property for a completely observed time series of a 
sufficiently isolated system, lost in the quantum domain, is 
recovered in a thermodynamic limit of a large number of particles.

We now show in detail that \gzit{e1} is inconsistent with quantum 
mechanics. According to the standard interpretation of probabilities as 
expectations of projection operators we have $P_{ab}=\<a|\Pi_b|a\>$
and $P_{bc}=\<b|\Pi_c|b\>$, where $\Pi_b=|b\>\<b|$ is the projector
to state $b$ (and similarly for $c$). In order that $P_{abc}$ makes 
sense in the standard interpretation we need $P_{abc}=\<a|\Pi_{bc}|a\>$ with a projector
$\Pi_{bc}$ expressing the proposition  $\Pi_b \& \Pi_c$ that $B$ is in
state $b$ and $C$ is in state $c$. 

In orthodox quantum logics ({\sc Birkhoff \& von Neumann} \cite{BirN},
{\sc Svozil} \cite{Svo}), one considers a projector as representing 
the subspace of eigenstates to the eigenvalue 1 of the projector and 
takes the logical operation $\&$  as subspace intersection. 
One finds that, whenever $\Pi_b \ne \Pi_c$, we have 
$\Pi_b \& \Pi_c = 0$ and thus $P_{abc}=0$. This is different from 
\gzit{e1}, proving that \gzit{e1} has nothing to do with quantum 
mechanics. And when $\Pi_b = \Pi_c$, we have $\Pi_b \& \Pi_c = \Pi_c$, 
hence $P_{abc}=\<a|\Pi_c|a\>=P_{ac}$, and \gzit{e2} turns out to be 
{\em always} true in the quantum logic calculus, in contrast to what 
Feynman claimed.

On the other hand, the statement $P_{abc}=0$ if $\Pi_b \ne \Pi_c$
seems somewhat counterintuitive, since if $\Pi_b\approx \Pi_c$ one 
might want to have $P_{abc}\approx P_{ac}$. In view of the fact that,
unless $\Pi_b$ and $\Pi_c$ commute, there is no simple {\em algebraic} 
expression relating $\Pi_{bc}$ to $\Pi_b$ and $\Pi_c$, 
one is not compelled to adhere to the rigid assignment 
$\Pi_b \& \Pi_c = 0$ and might prefer to
consider $\Pi_b \& \Pi_c$ to be undefined if $\Pi_b$ and $\Pi_c$ 
do not commute, interpreting this as an indication that $P_{abc}$ 
cannot be predicted (in principle) by the formalism of quantum theory 
(and thus may have any value). 

In view of the fact that the counterintuitive abundance of zero
projectors has further undesirable consequences to be discussed in
Section \ref{path}, this would seem a more satisfactory solution of 
Feynman's argument. A new quantum logic consistent with this 
interpretation is discussed in {\sc Neumaier} \cite{Neu1}.

\bigskip
%%%%%%%%%%%%%%%%%%%%%%%%%%%%%%%%%%%%%%%%%%%%%%%%%%%%%%%%%%%%%%%%%%%%%%%%
\section{Wigner's claim} \label{wigner}

Similar discussions as those given by Feynman usually suffer from the 
same problem that the claimed probability cannot be written as the 
expectation of a quantum proposition, i.e., a projector. For example, 
in his 1976 lecture notes on the `interpretation of quantum mechanics',
{\sc Wigner} \cite[p.288]{Wig} writes,

{\em
[...] it remains essentially correct to say that the basic statement of quantum mechanics can be given in a formula as simple as \gzit{e42}.
}

In adapted notation, for a finite sequence $a,b,c,d\dots$ of results 
of measurements of $A,B,C,D,\dots$, it reads
\setcounter{equation}{41} 
\lbeq{e42}
\begin{array}{ll}
P_{abcd...}&=\tr\Pi_a\Pi_b\Pi_c\Pi_d \cdots \Pi_d\Pi_c\Pi_b / \tr\Pi_a\\
           &=\<a|\Pi_b\Pi_c\Pi_d \cdots \Pi_d\Pi_c\Pi_b|a\>
            ~~~\mbox{if $a$ is a pure state.}
\end{array}
\eeq
\setcounter{equation}{5} 

Now there is a problem with \gzit{e42}. It implies that repeating
the measurement of $A$ immediately after a first measurement of $A$ 
gives the result of the first measurement with certainty. This is
possible only if we assume that the measurements performed are ideal.

But (cf. Wigner's discussion in pp.283-284 of \cite{Wig}; see also 
\cite{JooZ}) one cannot 
make an ideal measurement of an observable with a continuous spectrum;
and the quantum mechanical analysis of the measurement process shows 
that an ideal measurement of any quantity takes, strictly speaking, an 
infinite time. And even allowing for that, the
only measurable observables would be the functions of scattering
invariants (p.298). Hence formula \gzit{e42} can 
apply to real situations only approximately.

Therefore, unlike Schroedinger's equation, \gzit{e42}
cannot be considered a `basic statement' and must be banned from 
the foundations. It should rather be a consequence of more elementary, 
exact features of the theory.

\bigskip
Indeed, Wigner `derives' \gzit{e42} from first principles. 
The argument, given on pp.286-287, is identical to Feynman's, 
assuming (without comment) the Markov property that provides the 
independence that allows to multiply probabilities:

{\em
If the first observation [...] This probability is 
$|(a_\kappa,b_\lambda)|^2$. The probability that the next measurement  
[...] is then $|(b_\lambda,c_\mu)|^2$ and the probability of both 
outcomes [...] is $|(a_\kappa,b_\lambda)(b_\lambda,c_\mu)|^2$ [...].
}

Thus the formula, based on invalid reasoning, is suspect. 

This suspicion is confirmed by realizing that the operator whose 
expectation is taken is generally not a projector, 
hence does not correspond to a `proposition' in the 
traditional quantum logic calculus. But what sense should it make to
talk about the probability of a statement that is not even logically
well-formed? 

It seems to me that the only formula for probabilities verified
(abundantly) by experiments is the formula 
(see, e.g., equation (1.8) in Chapter 3 of {\sc Davies} \cite{Dav})
\lbeq{e6a}
P(B \in E|\rho)=\tr \rho B(E)
\eeq
for the probability that, in a mixed state with the density matrix 
$\rho$, the generalized observable (POV measure) $B$ has a value 
from a set $E$. For ordinary observables $B$ and singleton 
sets $E=\{b\}$, the `effect' $B(E)$ reduces to a projector $\Pi_b$. 
Equation \gzit{e6a} contains a special case of \gzit{e42} only,
\lbeq{e6}
\begin{array}{ll}
P_{ab}&=\tr\Pi_a\Pi_b / \tr\Pi_a \\
      &= \<a|\Pi_b|a\>= |\<a|b\>|^2
       ~~~\mbox{if }\Pi_a=|a\>\<a|,~\<a|a\>=1.
\end{array}
\eeq 

It is interesting to note that Wigner's formula \gzit{e42} reappears 
as the basic formula in the theories of consistent
histories (e.g., equation (4.3) in {\sc Omn{\`e}s} \cite{Omn}).
However, as mentioned on p.143 of \cite{Omn}, only the case of two
reference times can be proved, and this is equivalent to \gzit{e6a}
(with $\rho$ replaced by $E_1\rho E_1$, the mixed state obtained after 
the first state reduction).

\bigskip
However, we can give the probability \gzit{e42} an alternative, 
correct meaning: If $a$ is a pure state then \gzit{e6a} implies that
$P_{abcd}$ is the probability of measuring 
$\Pi_a\Pi_b\Pi_cD\Pi_c\Pi_b\Pi_a$ in that state! 
In general, therefore,  Wigner's formula gives the  probability 
of measuring in state $a$ the last element in a finite sequence $A$, 
$\Pi_aB\Pi_a$, $\Pi_a\Pi_bC\Pi_b\Pi_a$, 
$\Pi_a\Pi_b\Pi_cD\Pi_c\Pi_b\Pi_a$, $\dots$, assuming that the 
measurements resulted in reductions of the intermediate states to the 
states represented by the projectors $\Pi_a, \Pi_b, \Pi_c, \dots$.

To see how we can possibly interpret this in terms of the 
uncontroversial part of quantum mechanics, we consider a quantized 
relative of Laplace's demon. 
Suppose there were a quantum demon with the unusual capacity to `see' 
every detail of a closed system, without interacting with the system.
(The demon doesn't need to be physically realizable; this is just a 
fictional argument to make a vivid point.) 
The demon leaves the whole system (consisting of the measured system 
together with the measuring device) undisturbed and only interprets our 
claims on measurements and compares it with the dynamics it sees.

The demon notes the time intervals $t,u,v,\dots$ between the 
successive measurements $A,B,C,D,\dots$. What does it see us measure? 
In the Heisenberg picture, with unlabelled operators at the time of the 
first measurement, the state of the complete system remains unchanged, 
and the measurements are those of $A,B(t),C(t+u),D(t+u+v),\dots$.
If we use the Heisenberg dynamics and introduce the commuting operators 
$\Phi_a=\exp(itH/\hbar)$, $\Phi_b=\exp(iuH/\hbar)$, 
$\Phi_c=\exp(ivH/\hbar)$, we see that the demon sees us measure
$A, \Phi_a^*B\Phi_a, \Phi_a^*\Phi_b^*C\Phi_b\Phi_a, 
\Phi_a^*\Phi_b^*\Phi_c^*D\Phi_c\Phi_b\Phi_a, \dots$.
(This is quite different from what Wigner asserts in his equation (43), 
where he assumes an {\em additional} Heisenberg dynamics of the 
operators between the state-reducing measurements.)

Comparing with the above interpretation of \gzit{e42}, and noting that
projectors are Hermitian, we see that the difference between the
unproved formula \gzit{e42} and the demon's objective description of
our sequence of measurements is the replacement of the commuting unitary
$\Phi$'s in the complete system by the noncommuting idempotent 
Hermitian $\Pi$'s in the measured system alone. 

Thus \gzit{e42} is valid precisely to the extent that the time-dependent
operators (in the Heisenberg picture), as calculated by the Heisenberg 
equation for the {\em complete} experimental setup at the time the 
next measurement is taken, can be approximated in the assumed 
(time-independent) state of the investigated {\em subsystem} by a 
two-sided multiplication with these projectors. 

This shows clearly {\em the aim of a correct measurement theory:} 
It must exhibit broad sufficient conditions for a system that guarantee,
for all observables to be measured, the validity of this approximation, 
and hence a complete reduction of the wave packet.

I'd like to challenge the adherents of \gzit{e42} to devise and 
perform experiments testing validity, accuracy and limits of 
\gzit{e42} when more than two (noncommuting) projectors 
are involved and the measurements are not ideal. The findings should 
coincide with the interpretation in the Heisenberg picture just given. 
(For the realization of arbitrary discrete projectors, note that these 
can always be brought into the form $UDU^*$ with diagonal projectors 
$D$ and unitary $U$; hence {\sc Reck} et al. \cite{RecZB} applies.)

On the basis of such experiments it might also be possible to decide
whether it is at all possible to perform experiments for  
measurements of probabilities that cannot be expressed as the 
expectation of a projector.

\bigskip
%%%%%%%%%%%%%%%%%%%%%%%%%%%%%%%%%%%%%%%%%%%%%%%%%%%%%%%%%%%%%%%%%%%%%%%%
\section{Schr{\"o}dinger's argument} \label{schroe}

In his 1936 paper 
``Die gegenw{\"a}rtige Situation in der Quantenmechanik''
\cite{Sch} (where the famous cat paradox appears), 
Erwin {\sc Schr{\"o}dinger} gives the following argument 
(pp. 156-157 of the English translation): 

{\em
At first thought one might well attempt likewise to refer back the 
always uncertain statements of Q.M. to an ideal ensemble of states, 
of which a quite specific one applies in any concrete instance -- 
but one does not know which one. That this won't work is shown by the 
one example of angular momentum, as one of many. 

Imagine [...] the 
point $M$ to be situated at various positions relative to $O$ and 
fitted with various momentum vectors, and all these possibilities 
combined into an ideal ensemble. Then one can indeed choose these 
positions and vectors that in every case the product of vector length 
by length of normal $OF$ [where $F$ is the point closest to $O$ on the 
line through $M$ along the momentum vector] yields one or the other of 
the acceptable values -- relative to the particular point $O$. But for 
an arbitrary different point $O'$, of course, unacceptable values occur.
[...] 

One could go on indefinitely with more examples.
[...] Already for the single instant things go wrong. At no moment 
does there exist an ensemble of classical states of the model that 
squares with the totality of quantum mechanical statements of this 
moment. [...] we saw that it is not possible smoothly to take over 
models and to ascribe, to the momentarily unknown or not exactly known 
variables, nonetheless determinate values, that we simply don't know.
}

The problem is indeed unsolvable if one insists on the existence of a 
{\em single} vector $J$ of values determining all linear combinations 
$u\cdot J$ of the angular momentum. For example, in measuring the 
inner product $u\cdot J$ of the angular momentum $J$ with four or more 
triplewise linearly independent unit vectors $u$, one obtains in each 
case values from the discrete 
spectrum of $u\cdot J$, and these spectra are inconsistent with a 
precise value of $J$. This is explained by Schr{\"o}dinger in detail 
on p. 164 in 
the context of a simpler one-parameter example involving the 
observables $p^2+a^2q^2$ where $a$ is a parameter. Commenting on 
it, Schr{\"o}dinger writes on p. 165:

{\em
Should one now think that because we are so ignorant about the 
relations among the variable-values held ready in one system, that 
none exists, that far-ranging arbitrary combination can occur? 
That would mean that such a system of "one degree of freedom" would 
need not merely two numbers for adequately describing it, as in 
classical mechanics, but rather many more, perhaps infinitely many.
}

Schr{\"o}dinger dismisses this as not viable, but as I'll show 
now, there is a serious possibility that precisely this is the case,
thus invalidating Schr{\"o}dinger's conclusion that it is not possible 
to ascribe to the momentarily unknown or not exactly known variables 
determinate values. His conclusion is only valid if one ascribes
to each observable vector $K$ a determinate value that is to be used in the calculations
for {\em all} $u\cdot K$. 

However, as was observed already by {\sc von Neumann} 
\cite[IV.1.E]{vNeu}, one cannot, in general, combine the measuring 
recipes for two noncommuting observables to one for their sum, so that
the sum of two observables is only implicitly characterized through 
the axioms. This should forbid the naive use of values for the
components of $K$, say, to calculate the values of $u\cdot K$.

As I shall show now, one may consider each $u\cdot K$ as a 
(not necessarily classical) random variable in its own right, and 
determine their relationship 
for different $u$ not by ordinary algebra but by the statistics derived
from the standard quantum mechanical recipes.

For simplicity, I'll take for $K$ in place of the angular momentum $J$ 
the Pauli spin vector 
$\sigma$ which shows precisely the same problems as $J$. For each 
unit vector $u\in \Rz^3$, the operator $u\cdot \sigma$ has the 
simple eigenvalues $1$ and $-1$; thus these are the possible 
values of $u\cdot \sigma$. The projector to an eigenstate of 
$u\cdot \sigma$ corresponding to the eigenvalue $1$
is $\Pi_u:= \half(1+u\cdot \sigma)$, with $\tr \Pi_u=1$, and the 
corresponding projector for the eigenvalue $-1$ is simply
$\Pi_{-u}$. For any unit vector $v\in \Rz^3$, the probability for 
$v\cdot \sigma$ having the 
value $1$ when  $u\cdot \sigma$ has the value $1$ is then, 
according to \gzit{e6}, 
\lbeq{e7}
\begin{array}{ll}
P_{uv}&=\tr \Pi_u\Pi_v = \tr (1+u\cdot \sigma)(1+v\cdot \sigma)/4\\
      &=\tr(1+(u+v)\cdot\sigma+u\cdot v+i(u \times v)\cdot\sigma)/4\\
      &=\half(1+u\cdot v),
\end{array}
\eeq
and the probability for $v\cdot \sigma$ having the value $-1$ when 
$u\cdot\sigma$ has the value $1$  is therefore 
$P_{u,-v}= \half(1-u\cdot v)$, adding up to $1$, as it should be.
Moreover, $P_{u,-u}=0$, as it should be. The probabilities depend 
continuously on $u$ and $v$, in a very natural way. 

One would get 
precisely the same probabilities if one had extremely fragile,
classical spheres, painted white on one hemisphere and black on the 
other one, and each sphere would be destroyed after observing the
color at a single point. One could still calculate probabilities for
the colors of the other points, and any rotationally symmetric 
classical probability model produces precisely the same probabilities 
as we just calculated for the spin. 

More details on such a classical model for electron spin can be found
in {\sc Kochen \& Specker} \cite{KocS}. They show that, under natural
assumptions, such a classical model is restricted to two-state systems. 
(In a companion paper \cite{Neu1}, I'll show how one can reinterpret
quantum mechanics in such a way that the assumptions of Kochen and 
Specker become irrelevant, thus removing this very restictive 
conclusion.) In any case, one such model is enough to invalidate
the cogency of Schr{\"o}dinger's conclusion. 

\bigskip
To summarize,
Schr{\"o}dinger's argument demonstrates that, in quantum 
mechanics, one cannot calculate the values for linear combinations of
noncommuting observables from the values of the observables themselves;
but this is already obvious from the properties of the spectrum of 
operators. Only the values of functions of {\em commuting} observables
(and, in particular, of functions of a single observable) can be
predicted with certainty from the values of the observables themselves.

However, as was shown for the case of linear combinations of
the Pauli spin matrices, his argument does not demonstrate that one 
cannot assign natural values to all observables that lead to natural
and consistent classical probabilities. 

The nature of 
quantum observations may put severe limits on what is observable 
through experiment and how these are combined to estimate the values 
of other observables, but it does not seem to put restrictive limits 
on the joint values of noncommuting observables. Indeed, 
{\sc von Neumann} \cite{vNeu} shows in the same section mentioned above
how to get estimates of joint probabilities of noncommuting 
quantities if sufficiently large ensembles are available.

\bigskip
%%%%%%%%%%%%%%%%%%%%%%%%%%%%%%%%%%%%%%%%%%%%%%%%%%%%%%%%%%%%%%%%%%%%%%%%
\section{Particle paths} \label{path}

It is widely believed that it is impossible to ascribe definite paths
to moving quantum particles, except at special points where the position
has been measured.
The particle paths seen in a cloud chamber, say, can still be explained
as an illusion created by correlation patterns among the atoms ionized 
by the particles; see {\sc Mott} \cite{Mot}. No such analysis seems to 
be available, however, that would explain why we see macroscopic 
bound states (such as you or me) move along fairly definite paths.

The traditional context for discussing the inconsistency of definite
particle paths is the double slit experiment ({\sc Bohr} \cite{Boh}),
where it is claimed that electrons
(or photons) passing a diaphragm containing the two slits cannot
be said to have passed through one or the other slit, or -- even worse 
-- are said to have passed through both, in a way (superposition) that 
defies intuition. 

Unfortunately, the discussion seems always to be connected to
measurement questions, so that it is difficult to discern the 
mathematical content of the arguments that claim {\em nonexistence}
of the path (in contrast to its {\em limited measurability} only,
{\sc Wootters \& Zurek} \cite{WooZ}). However, I gathered some 
indirect evidence about particle paths by reading between the lines 
of some papers.

{\sc Feynman} \cite{Fey}, in laying ground work for his 
(nonrelativistic) path integral formalism, discusses on p. 371 the 
probability of particle paths as his first postulate:

{\em
If an ideal measurement is performed to determine whether a particle 
has a path lying in a region of space-time, then the probability that 
the result will be affirmative is the absolute square of a sum of 
complex contributions, one from each path in the region.
} 

This postulate is used to show (among other things) that the most 
likely paths are those close to the path determined by the least 
action principle.

No comment is given on the precise mathematical meaning of the terms 
involved (except for the term ``each''); thus we need to see whether
we can interpret this in an orthodox way. In the spirit of the standard
interpretation of quantum mechanics, we need to assign to each
(let us say open and bounded) region $\Omega$ of space-time a 
projector $\Pi_\Omega$ to the subspace $\Hz_\Omega$ of all wave 
functions belonging
to states where the particle is with certainty in $\Omega$.

Unfortunately, particle positions at different times are 
represented by operators $x(t)=\exp(iHt)x(0)\exp(-iHt)$ that generally 
do not commute with each other, with the consequence that it is unlikely
that they have common eigenvectors, which would be the elements of 
$\Hz_\Omega$. Thus, it is very likely that, for most $\Omega$, this 
subspace consists of zero only; and $\Pi_\Omega$ would vanish. 
This casts serious doubt on the applicability of the quantum logic 
recipe to defining the projector $\Pi_\Omega$. 

Perhaps it is even impossible to make consistent assignments of all 
projectors $\Pi_\Omega$, in view of the fact that alternative attempts 
in consistent histories interpretations run into contradictions 
({\sc Kent} \cite{Ken}). (But possibly these contradictions are due
to the non-projector nature of these histories; cf. {\sc Isham}
\cite{Ish}.)

However, we can be more modest and only look at special sets of paths 
for which we may define proper projectors. For example, we may define 
the distance of two paths $\xi^k:[\alpha,\omega]\to \Rz^3$ ($k=1,2$) by
\lbeq{e8a}
d(\xi^1,\xi^2)=\sqrt{\int_\alpha^\omega (\xi^1(t)-\xi^2(t))^2dt}.
\eeq
Then, for each specific path $\xi$, we can use the position operators
$x(t)=\exp(iHt)x(0)\exp(-iHt)$ to define the observable
$\Delta_\xi:=d(x,\xi)$ that measures the distance of the particle path 
from $\xi$. Suitable projectors are now the projectors $\Pi_{\xi,\eps}$
projecting to the subspace spanned by the wave functions corresponding 
to states in which $\Delta_\xi$ has with certainty a value not
exceeding $\eps$. Then, for a system in an arbitrary pure state $a$,
\lbeq{e8}
P(d(x,\xi)\le \eps)=\<a|\Pi_{\xi,\eps}|a\>
\eeq
is, according to the most orthodox interpretation of quantum mechanics,
a well-defined quantum mechanical probability for the event that 
the particle path is, in the root mean square sense, within $\eps$ of 
the path $\xi$. Note that, in principle, this formula can be tested 
experimentally; and by varying $\eps$, the complete distribution of 
$d(x,\xi)$ can be found. The natural path to be used for $\xi$ is the 
expectation 
\lbeq{e9}
\xi(t)=\<a|x(t)|a\>
\eeq
or a computable approximation to it (such as the classical path of the 
particle).

With formulas such as \gzit{e8a}--\gzit{e9} available, it would be 
completely unreasonable to assume that a particle has no definite
path. If it hadn't, what should be the meaning of its expectation
\gzit{e9} and of the probability \gzit{e8} of it being close to a 
given path!?

\bigskip
It remains to discuss how this finding can be reconciled with the 
double slit experiment. In an implementation of von Neumann's method 
mentioned in Section \ref{schroe} to measure probabilities for two 
noncommuting observables, {\sc Wootters \& Zurek} \cite{WooZ} 
measure in a modified double-slit experiment both position and 
momentum of particles passing a slit:

{\em 
If the measured momentum is positive, then we will guess that the 
photon passed through slit $A$; if it is negative, then we will guess 
that the photon passed through slit $B$. Clearly some of our guesses 
will be wrong -- there are photons that have positive values of 
measured momentum even though their actual momentum was negative and 
they went through slit $B$. 

[...] 
these two measurements cannot both be performed for the same photons, 
and so [the figures] cannot refer to the same experiment. Hence, in 
accord with the Copenhagen interpretation of quantum mechanics, there 
is no paradox. The complementarity principle does not prevent photons 
from behaving once as waves and once as particles. It only states that 
the same photon should not reveal this "split personality" in the same 
experiment. [...] 

Despite the fact that we know with $99\%$ certainty 
the paths of the photons, they still have strong wavelike properties.
[...] we have presented a result which, although not paradoxical, was 
nevertheless surprising (that is, that one can make a fairly precise 
determination of the slit through which each photon comes with only a 
slight disturbance of the interference pattern).
}

The finding loses its surprise in the light of the above considerations
that suggest existence of the path but with intrinsic limitations on 
its observability.

For the original double-slit experiment, the arguments generally agree 
on the observation that there is no way to predict through which slit a 
particle {\em will pass}. However, it is possible to compute with high 
confidence through which a particle {\em has passed} when it has been 
observed at the photographic plate. To do so, one only needs to compute 
the probabilities \gzit{e8} for two typical paths $\xi$ connecting each
slit with the position where the particle was absorbed by the 
photographic plate. By comparing the likelihood ratios at a given 
significance level one will find an exceedingly high likelihood for 
most particles having passed through the slit closer to their recorded 
position on the photographic plate. 

Thus, after the event, enough information is available to decide 
reasonably reliably on the particle path. 
One could also calculate the values of the actions of the
paths through the two slits ending at the observed position, and
use the stationary phase approximation of the Feynman path integral to 
get (approximations to) the required probabilities. 
In any case, the statistical analysis is essentially 
similar to that used for all experiments where an event must be 
reconstructed from indirect measurable information and from the way 
the experiment was prepared.

\bigskip
%%%%%%%%%%%%%%%%%%%%%%%%%%%%%%%%%%%%%%%%%%%%%%%%%%%%%%%%%%%%%%%%%%%%%%%%
\section{Conclusion} \label{conc}

The strongest arguments 
against realism available in the literature dissolved under a 
scrutinized analysis, using only that part of quantum mechanics for 
which there is almost universal agreement about its validity.

This opens the door for an interpretation that, while respecting 
the indeterministic nature of quantum mechanics, allows to speak of 
definite but only partially measurable values for all observables at 
any time. This kind of realism is consistent with the intrinsic 
indeterminism 
required by Heisenberg's uncertainty relation if we distinguish 
carefully between {\em what is} and {\em what is measurable}.

Though not proved by the present investigations, it appears that, 
independent of the detailed description, 
a cautious realistic interpretation of quantum mechanics is in 
full accord with the generally accepted quantum mechanical formalism.

Here {\em realistic} means that in a completely specified state
(not to be confused with the `pure states' of quantum mechanics) 
{\em all} observables have definite values at all times that are, 
however, only partially measurable, according to the stochastic 
predictions of quantum mechanics.

And {\em cautious} means, that one has to take into account the 
following four restrictions:

1. Only probabilities defined by orthogonal projectors (for sharp
observables) or POV measures (for smeared observables) via equation
\gzit{e6a} are measurable and predictable.

2. For noncommuting projectors, the logical operation `and' is 
not defined; the corresponding questions may be asked (and may have
definite, i.e., logically consistent answers) but a precise answer 
cannot be found by experiments.

3. From the values of commuting observables one can deduce values of 
functions of these observables, but from the values of noncommuting 
observables one cannot even deduce values of their linear combinations. 
Instead, functions of noncommuting observables must be considered as 
random variables on their own whose expectation values (and probability distributions) must be calculated from the general quantum mechanical 
formalism and not from a classical inference of measured linear 
combinations.

4. Due to quantum inseparability, the Markov property for a completely 
observed time series is lost.

A constructive proposal for such an interpretation will be given
in a companion paper \cite{Neu1}.

\bigskip
%%%%%%%%%%%%%%%%%%%%%%%%%%%%%%%%%%%%%%%%%%%%%%%%%%%%%%%%%%%%%%%%%%%%%%%%
\section*{Appendix: Some final speculations} \label{spec}

In this section I want to be more speculative, at the risk of being less
precise, less cogent, and more vulnerable about some issues discussed. 
I shall mention some ideas and `conclusions' that are related to the 
preceding analysis though far from being consequences of the rigorous
observations of the preceding sections.

A realistic interpretation of quantum mechanics is independent of the 
measurement problem and in better accord with classical (say, 19th
century) intuition. 
As we have seen, there seem to be no longer strong arguments against a 
cautious realistic interpretation of quantum mechanics. 

A problem reamining, and one that 
obscured for a long time the underlying simplicity, is the lack of 
observable information on the quantum level. As {\sc Schr{\"o}dinger}
\cite{Sch} puts it (p.159 of the English translation), referring to
the complementarity between position and momentum, 

{\em
[...] the momentary statement content of the $\psi$-function is far 
from complete; it comprises only about $50$ percent of a complete 
description.
}

However, unlike Schr{\"o}dinger, I conclude that, while the remaining 
50 percent will probably always remain unobservable, they may have a 
reality just as objective as the (ideally) observable 50 percent.
The fact that in many cases we can choose which 50 percent of a 
complete description we want to observe underlines this conclusion.

The fact that both past and future boundary conditions are needed 
(and sufficient) to 
locate a particle path is consistent with the observation in Section 
\ref{feynman} that states anticipate something about their whole future.
It is also reminiscent of the action-at-a-distance 
formulation of classical electrodynamics by {\sc Wheeler \& Feynman} 
\cite{WheF,WheF2}, where electromagnetic radiation is explained through
interaction with particles in the future. 

By observing the right $50\%$, namely past {\em and} future position 
boundary conditions about {\em one} of a pair of conjugate variables at 
both boundaries, we might be able, in principle, to reconstruct 
the complete intermediate picture as reliably as in the classical case,
where one has the same option. But in the classical case one has the 
additional and, for us as subjects acting based on past information 
only, more useful option to predict the particle path from initial 
conditions only (position and momentum in the past).

This state of affairs can be summarized in the statement:

{\em 
Physics essentially describes nature as if everything had already 
happened, and then expresses its laws as information about 
observed correlations. Since some of the correlations involve time, 
it is possible to partially predict the future from the past, or
the past from the future, or an intermediate situation from past and
future observations.
}

This summary also explains neatly why questions such as the flow of 
time or free will cannot be discussed within the framework of physics. 
Whether or not time flows, whether or not our will is free, the 
four-dimensional picture resulting from the course of nature, whether 
or not influenced by us, can (in a gedankenexperiment) be replayed,
after everything has happened, like a movie. In the replay,
everything is determined, and there is only the illusion of free will,
just as we are used from the cinema. But the physics, expressed in 
the correlations between the parts of the movie, is identical to that 
in the original version. 

The principle of physics, that it restricts attention to that which 
remains the same after everything happened, makes physics very powerful 
in that it allows us to investigate a past of billions of years and to 
anticipate a future of billions of years. But the same principle also 
generates its intrinsic limitations, that physics must be silent about 
everything that cannot be captured by this static, four-dimensional 
view of nature. This includes both questions such as ``how does 
reality happen?'', and many of the subjects of most interest to 
people: freedom, purpose, and consciousness.

%%%%%%%%%%%%%%%%%%%%%%%%%%%%%%%%%%%%%%%%%%%%%%%%%%%%%%%%%%%%%%%%%%%%%%%%

\end {document}